\documentclass[10pt,journal]{IEEEtran}

\usepackage{times,amsmath,epsfig,psfrag,subfigure,multirow}

\begin{document}
%
% paper title
% can use linebreaks \\ within to get better formatting as desired
\title{Varactor-Based Dynamic Load Modulation of High Power Amplifiers}

\author{Ali Soltani Tehrani, Hossein Mashad Nemati, Haiying Cao,  Thomas Eriksson, and Christian Fager% <-this % stops a space
\thanks{This research has been carried out in the GigaHertz Centre in a joint research project financed by the Swedish Governmental Agency of Innovation Systems (VINNOVA), Chalmers University of Technology, Ericsson AB, Infineon Technologies and NXP Semiconductors.}
\thanks{A. Soltani Tehrani and T. Eriksson are with the Department
of Signals and Systems, Chalmers University of Technology,
Gothenburg, Sweden e-mail: [asoltani,thomase]@chalmers.se.}% <-this % stops a space
\thanks{C. Fager is with the Department of Microtechnology and Nanoscience, Chalmers University of Technology, Gothenburg, Sweden e-mail: christian.fager@chalmers.se.} 
\thanks{H. Cao, and H.M. Nemati are with Ericsson AB [haiying.cao,hossein.nemati]@ericsson.com}}

% make the title area
\maketitle

\begin{abstract}
%\boldmath
In this work, dynamic load modulation of high power amplifiers using a varactor-based
tunable matching network is presented. The feasibility of dynamic tuning and
efficiency enhancement of this technique is demonstrated using a modular design
approach for two existing high efficiency power amplifiers (PA), a 7-W class-E, and a 10-W class-J power amplifier PA at 1 GHz. For this purpose and for each of the PAs, a simple quasi-static inverse model is developed allowing an efficiency-optimized control of the PA and the varactor-based tunable matching network. Modulated measurements using a single carrier WCDMA signal with 11.3 dB peak-to-average ratio (PAR) indicate about 10 to 14 percentage units improvements in the average power-added efficiency (PAE) for the complete architecture.

\end{abstract}

\IEEEpeerreviewmaketitle

\section{Introduction}
\IEEEPARstart{P}{ower} amplifiers (PAs) are vital components of transmitter architectures that convert the supplied DC power to information carrying power at radio frequency (RF). PAs are normally designed to have high peak-power efficiency, so the conversion has as little loss as possible. High power efficiency is important in mobile applications, where the power is normally drawn from a limited power supply, but it is also important in base stations, where it can substantially reduce the amount of consumed energy.

With the need to fully utilize the limited bandwidth spectrum, modern wireless communication systems utilize variable envelope modulation schemes. A common property of these schemes is a high peak-to-average ratio (PAR) of the communication signal. The strict linearity constraints of the communication systems on the PA result in the need to operate the PA at a significant output power back-off. This severely degrades the average power efficiency for even PAs with high peak efficiency.

Since the power amplifier is the main power-consuming device in the transmitter architecture, improving the average power efficiency of the PA has received a considerable amount of attention in the literature. Many techniques have been proposed to enhance the power efficiency of PAs in back-off operation, and methods such as dynamic biasing \cite{yang}, dynamic supply modulation (DSM) \cite{wang} -- also known as envelope tracking -- and dynamic load modulation (DLM) \cite{raab} have shown to be the most promising. DLM may be realized by either varactor based tunable matching networks \cite{fuMTT,hosseinMTT}, or by active devices as in Doherty amplifiers \cite{doherty,cripps}.

While the improvement in efficiency in back-off with dynamic biasing and dynamic supply modulation is impressive, the need for active circuits increases the overall power consumption. Doherty amplification has been shown to be effective \cite{pelk}, but the increase in cost, complexity and size and the need for additional transistor devices and combining circuits are some of the drawback of this architecture \cite{fuMTT}. Compared to these architectures, in order to control a varactor-based tunable matching network, no significant power is required. This has the added benefits that this architecture is better equipped to cope with wideband signals and that it generally results in simpler designs \cite{raab}.

\begin{figure}[t]
\centering \psfrag{1}[c][c][1]{$x$} \psfrag{2}[c][c][1]{$V_\text{c}$}
\psfrag{3}[c][c][1]{$y$}
\includegraphics[width=0.95\columnwidth]{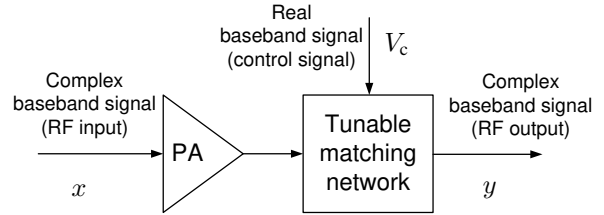}
\caption{The general transmitter architecture for dynamic load modulation.}
\label{architecture}
\end{figure}

A general block diagram of a dynamic load modulation transmitter architecture is shown in Fig.~\ref{architecture}. The output of a power amplifier is connected to an electrically tunable matching network. By co-controlling the RF input signal $x$ and the baseband control voltage $V_\text{c}$ to the matching network, the output signal $y$ can be manipulated in a way that either the average power efficiency is increased, the linearity is improved, or even both are improved simultaneously.

Varactor-based matching networks have been proposed for use as dynamic tuners \cite{neo,buisman}. A variety of network topologies have been explored to achieve the tunable matching that is suitable for DLM applications. In \cite{raab}, \cite{fuMTT} and \cite{fu1} a T-network was used as the tunable matching network. In \cite{neo}, different configurations were analyzed and implemented to obtain the tunable matching network. Recently in \cite{hosseinMTT} a varactor-based matching network (VMN) was designed for high power amplifier operations. In this design, the modular approach allowed the separate design of the VMN and the PA, which greatly simplified the design process. The benefits of this design is further analyzed in this work.

In \cite{fuMTT} and \cite{soltani} dynamic load modulation of high power amplifiers was demonstrated. In \cite{fuMTT}, the drain efficiency of a medium power amplifier with dynamic load modulation for a 3GPP WCDMA signal with 3.3 dB peak-to-average ratio was improved by 5 percentage units. In \cite{soltani}, with the modular design of PA and VMN from \cite{hosseinMTT}, modulated measurements showed a 12 percentage unit improvement in average power added efficiency (PAE) while maintaining linearity for an 11 dB PAR WCDMA-like signal with 384 kHz bandwidth.

In this work, we expand on the input signal design and inverse modeling in \cite{soltani}, provide dynamic load modulation results on two separate highly efficient PAs and show wideband measurement results. After showing the practicability of this technique, the usefulness of the modular approach to VMN design is also shown. In order to utilize the VMN network for both PAs, only an altercation in the input signal design and a slight hardware change is necessary.

The paper is organized as follows. Section II gives a background on previous work, and explains the important results that are needed further. In Section III, the behavioral modeling is explained, and the signal creation for the DLM architecture is discussed in detail. The measurement setup and the modulated measurement results are analyzed in Section IV and conclusions are drawn in Section V.

\section{Background}
In this section, some of the main results of \cite{hosseinMTT,soltani,hosseinEMW} which are needed for the discussion in this work are reviewed.

In designing a dynamic load modulation network, two approaches have been taken in the literature. The first is a co-design of the PA and the VMN \cite{fuMTT}, and the second is a modular design of the PA and VMN separately \cite{hosseinMTT}. There are advantages to both approaches, but the modular approach allows the use of the same VMN with different power amplifiers having similar operation frequency, peak output power and transistor technology. As an added bonus, the VMN can be designed for an available PA and improve the efficiency in an ad-hoc fashion. In this work, the modular approach is used.

In \cite{hosseinMTT}, the load modulation architecture was achieved with a varactor matching network, and specific design issues regarding the varactors were discussed in detail. In order to achieve high linearity and high power, an antiseries connection of a varactor stack was utilized. In \cite{hosseinEMW} load pull measurements were used to show that the maximum obtainable efficiency by controlling both the input and the load impedance to the PA can be improved by 20 percentage units in back-off operation.

In Fig.~\ref{architecture}, if a quasi-static relationship is assumed between the inputs $x, V_\text{c}$ and the output $y$, the output signal can be written as \cite{soltani}:
\begin{equation}
y = f_{\text{A}}\left(|x|,V_\text{c}\right)e^{-if_{\varphi}\left(|x|,\angle x,V_c \right)},
\label{y_x}
\end{equation}
where the RF input signal $x$ is assumed to consist of a time-varying amplitude $|x|$ and phase $\angle{x}$, $f_\text{A}\left(\cdot\right)$ is the AM-AM function, $f_\varphi\left(\cdot\right)$ is the AM-PM function, and the baseband control voltage signal consists of an amplitude $V_\text{c}$.

\begin{figure}
\centering
\includegraphics[width=0.8\columnwidth]{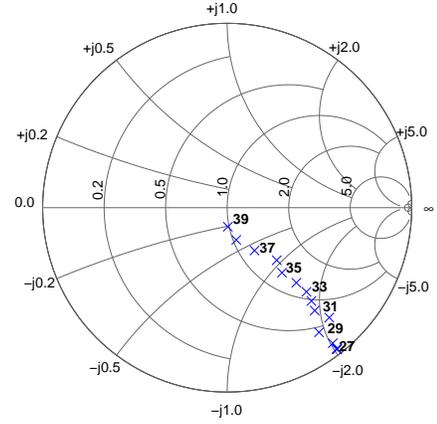}
\caption{Selected load impedances from the load pull measurements. The optimum load for each power level that maintains high efficiency is shown.}
\label{classEsmith}
\end{figure}

It must be taken into consideration that many combinations of $x$ and $V_\text{c}$ may exist to provide the same output signal. However, by constraining all such combinations to the ones that achieve the highest power efficiency, it is possible to construct a one-to-one relationship between the input signal and load impedance to the optimum output signal. This corresponds to a certain PA load impedance trajectory. The VMN design has to achieve this load-line as close as possible. The optimum trajectory to obtain high PAE for the PA used in this work is shown in Fig.~\ref{classEsmith} \cite{hosseinMTT}.

\begin{figure}
\centering
\psfrag{a}[c][c][1]{$x$} \psfrag{y}[c][c][1]{Varactor-based} \psfrag{x}[c][c][1]{Matching network} \psfrag{b}[c][c][1]{$V_c$} \psfrag{c}[c][c][1]{$y$} \psfrag{z}[c][c][1]{7W class E LDMOS PA}
\includegraphics[width=0.9\columnwidth]{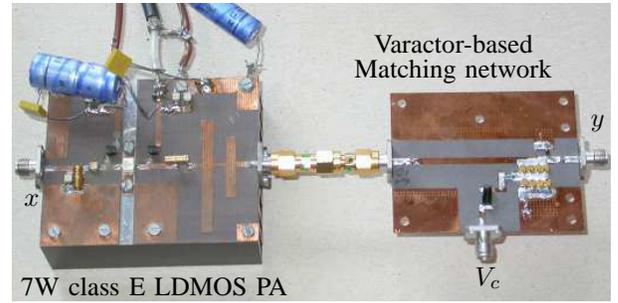}
\caption{Picture of the load modulation architecture for the class-E PA.}
\label{classE}
\end{figure}

In \cite{hosseinEMW}, an electrically controlled VMN is designed to approximate an efficiency optimized impedance for each output power level for a switched-mode class-E 7 W LDMOS PA operating at 1 GHz \cite{adahl}. This is also used in this work. The PA was biased at $V_g=3.5$ V and $V_d = 18$ V. The peak power added efficiency of the PA is around 60 percentage units. The corresponding architecture is shown in Fig.~\ref{classE}.

\begin{figure}
\centering
\includegraphics[width=0.9\columnwidth]{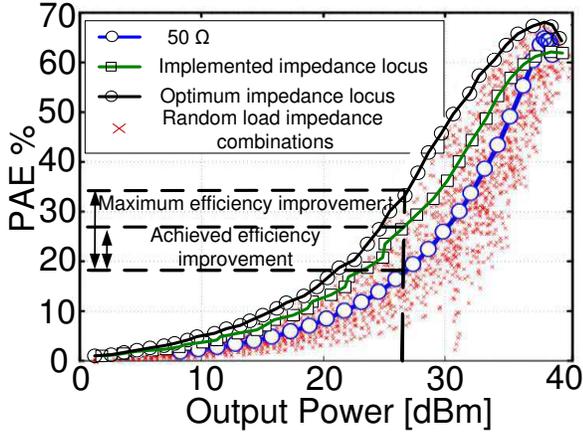}
\caption{Static performance of the dynamic load modulation architecture. The maximum achievable efficiency is shown to be 20\% while the implemented efficiency is around 11\% at 10 dB back-off, from \cite{hosseinMTT,soltani}.}
\label{back}
\end{figure}

The static measurements obtained by the constructed network from \cite{hosseinMTT} are shown in Fig.~\ref{back}. It can be observed that while the maximum efficiency enhancement was around $20$\% at 10 dB back-off, the obtained efficiency from the designed PA+VMN network is $11$\%. This is due to the losses and mismatch associated with the practical implementation \cite{hosseinMTT}.

\section{Signal Creation and Inverse Modeling}
The next step, after designing the necessary hardware, is to construct the optimum input signals. This process is explained in detail in this section.

\begin{figure}
\centering
\psfrag{a}[c][c][0.8]{$x$}\psfrag{b}[c][c][0.8]{$V_\text{c}$}\psfrag{c}[c][c][0.8]{$y$}
\subfigure[]{\label{fig:SISO}}\includegraphics[width=0.6\columnwidth]{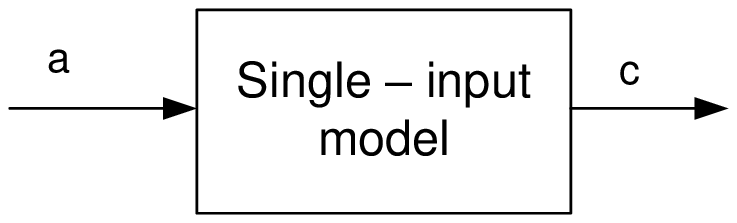}
\psfrag{a}[c][c][0.8]{$x$}
\subfigure[]{\label{fig:MISO}}\includegraphics[width=0.6\columnwidth]{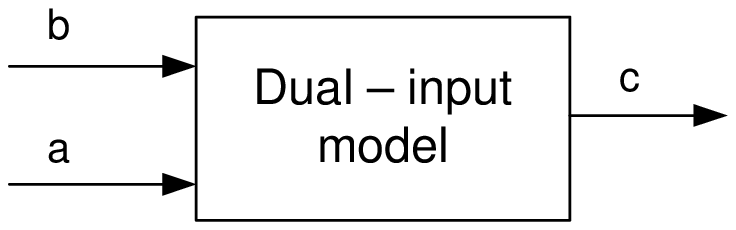}
\caption{Different transmitter architecture block diagrams, (a) is the common single input single output architecture and (b) is the block diagram for the dynamic load modulation and dynamic supply modulation architectures.}
\label{tranblock}
\end{figure}

From a signal processing perspective, the traditional transmitter architecture consisting of a sole power amplifier can be thought of as a single-input single-output system, as shown in Fig.~\ref{fig:SISO}. The normal definitions for efficiency and linearity are straightforward for this architecture. Transmitter architectures like DLM or envelope tracking have a dual-input nature, as shown in Fig.~\ref{fig:MISO}. Linearity for such systems is not well-defined. In order to be able to compare these two architectures, it is necessary to analyze linearity further.

\subsection{Extraction of dual input control functions}

A common interest for both architectures is that the output signal, $y$, be a linear representation of the communication signal, while maintaining a high efficiency for the PA. To compare the two architectures fairly, this final output signal $y$ can be used.

\begin{figure}
\centering \psfrag{2}[c][c][1]{$x$} \psfrag{3}[c][c][1]{$V_c$}
\psfrag{4}[c][c][1]{$y$}\psfrag{1}[c][c][1]{$u$}
\includegraphics[width=0.9\columnwidth]{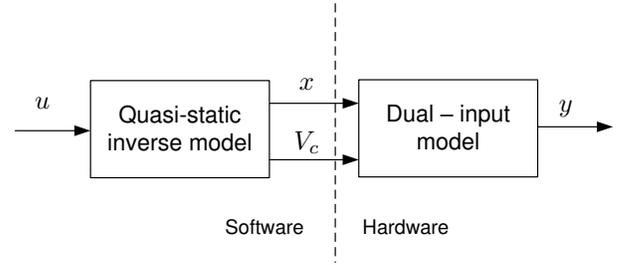}
\caption{The block diagram for input design for the DLM architecture.}
\label{modeling2}
\end{figure}

For the dual-input architecture, we have one extra degree of freedom to change the two inputs to achieve the desired output signal. Because of the dual input nature of the architecture, the design of an appropriate complex-valued RF PA input signal $x$, and the baseband control voltage $V_c$ for the matching network is challenging. In order to find the input signals, a simple static nonlinear model is used. The block diagram of dual-input architectures is shown in Fig.~\ref{modeling2}. By defining $u$ as the desired output signal, we can define the linearity of this architecture as the output signal $y$ vs the desired output signal $u$.

The input signals ($x$ and $V_\text{c}$) to the DLM architecture can be designed by the following procedure.

\begin{description}
\item[Step 1] The output signal $|y|$ of the PA + VMN is measured for different control voltage settings $V_c$ with varying drive levels $|x|$. The static results from the load-pull measurements with the Large Signal Network Analyzer (LSNA) from Section II are used. In this work, the input signal power was swept in 1 dB steps. The resulting PAE for the different input signal power levels and control voltages is shown in Fig.~\ref{allcomb}.

\begin{figure}
\centering
\includegraphics[width=\columnwidth]{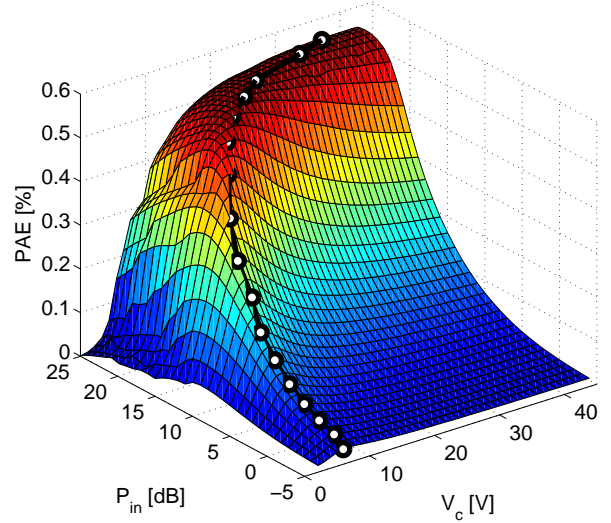}
\caption{PAE for the different combinations of PA input power and VMN control voltage. The black line represents the maximum obtainable PAE trajectory.}
\label{allcomb}
\end{figure}

\item[Step 2] For each output power level, a grid search is done to find the input signal power and control voltage that results in the highest PAE. In Fig.~\ref{allcomb}, this corresponds to the black line. By choosing the $V_\text{c}$ such that for each $P_\text{in}$ we are always on this black line, we can obtain the highest PAE versus output power for the PA.

\item[Step 3] Since the resulting values are discrete, for each output level, polynomial interpolation is used to find the corresponding combination of $V_c$ and input signal that corresponds to the highest efficiency. These functions are given in (\ref{aoutin1})-(\ref{aoutin3}). If necessary, extrapolation is also done.
    \begin{eqnarray}
    |x| =& f_{A_{\text{opt}}}\left(|u|\right)\label{aoutin1}\\
    \angle{x} =&
    f_{\varphi_{\text{opt}}}\left(|u|,\angle{u}\right)\label{aoutin2}\\
    V_{\text{c}} =& f_{Z_{\text{opt}}}\left(|u|\right)\label{aoutin3}
    \end{eqnarray}

    For this PA, the input-output transfer function for optimum PAE performance for equations (\ref{aoutin1}) and (\ref{aoutin3}) is shown in Fig.~\ref{out-inE} and for (\ref{aoutin2}) in Fig.~\ref{phaseE}.

    \begin{figure}
    \centering
    \psfrag{a}[c][c][1]{RF Input voltage $|x|$ [V]} \psfrag{b}[c][c][1]{RF Output voltage $|y|$ [V]} \psfrag{c}[c][c][1]{Control voltage $V_c$ [V]}
    \includegraphics[width=0.9\columnwidth]{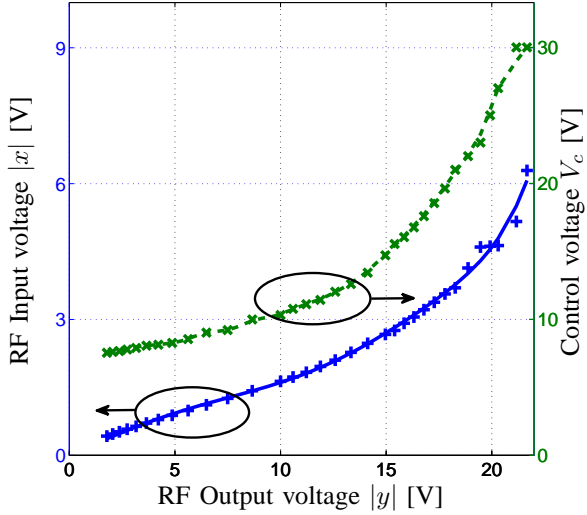}
    \caption{Output-input power relations for optimum PAE performance for the class-E PA. The blue crosses are the RF input voltages from static measurements and the blue line is the polynomial approximation. The green pluses are the control voltages from static measurements and the green dashed line is the polynomial approximation.}
    \label{out-inE}
    \end{figure}

    \begin{figure}
    \centering
    \includegraphics[width=0.9\columnwidth]{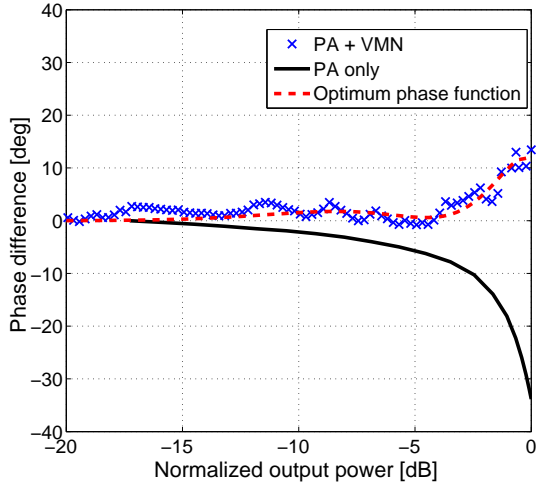}
    \caption{Insertion phase difference of the class-E power amplifier without
    a tunable network, with the VMN from static measurements, and the optimum phase transfer function , $f_{\phi_opt}$.} \label{phaseE}
    \end{figure}

\item[Step 4] By using the optimum functions (\ref{aoutin1})-(\ref{aoutin3}), the desired RF output signal is used as input to the functions and the corresponding complex-valued input $x$ and control voltage $V_\text{c}$ for each signal sample is found.
\end{description}

The optimum phase difference from the LSNA measurements for the architecture with/without the VMN is shown in Fig.~\ref{phaseE}. It is interesting to note that the PA + VMN combination actually results in lower AM/PM distortion than the PA itself. This can be understood since the purpose of the VMN is actually to cancel the reactive effects in the device and restore a purely resistive load to the intrinsic device source.

\subsection{Theoretical prediction values with modulated data}

Before testing the hardware, in this section we provide some predictions regarding the performance, to make an assessment on the potential of the technique. The desired output signal, $u$ that the DLM transmitter is tested with is a single carrier WCDMA signal with a PAPR of $11.3$ dB.

\begin{figure}
\centering
\psfrag{z}[c][c][1]{Probability distribution}
\includegraphics[width=0.85\columnwidth]{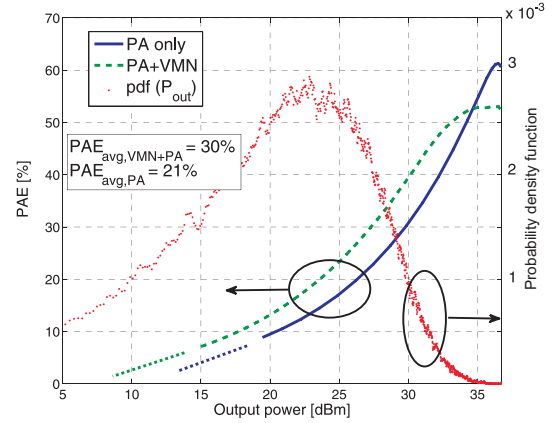}
\caption{Theoretical efficiency for the VMN network vs fixed $50~
\Omega$ load impedance for the class-E PA.} \label{pdfE}
\end{figure}

Fig.~\ref{pdfE} shows the prediction for the efficiency vs. output power for the class-E PA and the class E PA + VMN respectively. By applying the output power probability distribution function (pdf) of the WCDMA signal, the average efficiency can be calculated by averaging over this pdf.

The PAE averaged over the probability distribution of the signal for the PA without the varactor-based matching network is $21\%$, while the PAE with the
VMN is around $30\%$. The respective drain efficiency is $22\%$ and
$31\%$. Hence, with the use of the matching network, even though the peak
output power PAE is lower due to varactor losses, the overall predicted average efficiency improvement is around $9\%$.

\section{Modulated measurements}
In this section modulated measurements will be used to evaluate the performance of the dual input modeling presented.

\subsection{Measurement Setup}
\begin{figure}
\centering \psfrag{a}[c][c][1]{$x$}
\psfrag{b}[c][c][1]{$V_\text{c}$}
\psfrag{c}[c][c][1]{$\tilde{x},\tilde{V_\text{c}},\tilde{y}$}
\includegraphics[width=0.95\columnwidth]{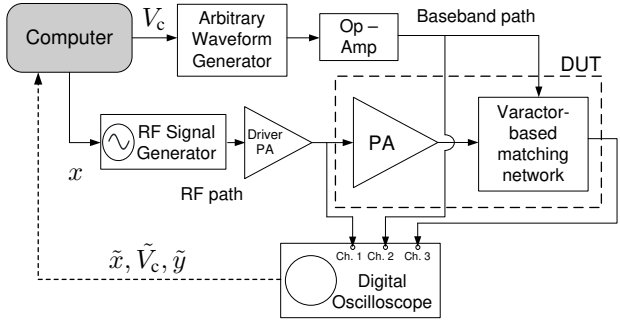}
\caption{Measurement setup used in the experiment.} \label{setup}
\end{figure}

The measurement setup used in this work is shown in Fig.~\ref{setup}, where $x$ is the complex-valued input signal (RF input) and $V_c$ is the real-valued control voltage. These signals and the output signal $y$ are measured by the oscilloscope and named $\tilde{x}$, $\tilde{V_c}$ and $\tilde{y}$ respectively.

After finding the optimum functions with the procedure from section III, the input signals are calculated and the complex baseband input signal is uploaded to an Agilent E4438C vector signal generator that acts as a modulator, and the control voltage for the VMN is uploaded to a Tabor Electronics, WW2572A arbitrary waveform generator (AWG). A four channel Agilent 54845A digital storage oscilloscope (DSO) is used in this case as a vector signal analyzer. This allows the simultaneous measurement of RF input and output signals, as well  as the varactor control voltage, see Fig.~\ref{setup}. The data is then downloaded to the  computer where downconversion to baseband and time alignment are done. All devices are connected by GPIB and triggered in synchronization. In order to increase the dynamic range  of the DSO, statistical averaging is done in the experiment  \cite{fager}. The results here  are based on 100 averaged measurements resulting in a dynamic range of approximately 45 dBc.

In order to achieve the necessary dynamic range for the VMN for this data, a
voltage swing of $6 - 27$ V on the varactors was needed (see Fig.~\ref{out-inE}). The AWG
used in this experiment could not provide such voltages, so a  simple
op-amp circuit was constructed to amplify the control voltage signal
to the necessary level.

An important issue for this experiment is the time alignment of the RF input signal to the matching network, and the baseband control signal. In the circuitry, different delays exist in the RF path and the baseband control signal path. If these signals are applied to the DLM transmitter architecture directly without a careful synchronization, it will result in unwanted distortion of the output signal. According to \cite{raab1} and \cite{rudolph}, even for a small time mismatch, distortion can be very strong. In order to achieve high efficiency and have better linearity, time alignment should be performed before uploading the signals to the measurement system. In this work, the time alignment was adjusted manually for minimum distortion of the output signal. More advanced algorithms to estimate the accurate delay can further improve the performance both in terms of linearity and power efficiency.

\subsection{Results -- Class E PA}
The results of the measurements on the DLM architecture are given in this section. In order to have a fair comparison with the PA-alone architecture, quasi-static predistortion was also done for the PA-alone results. This can be considered as a simple predistortion for both architectures.

\begin{figure}
\centering
\includegraphics[width=0.9\columnwidth]{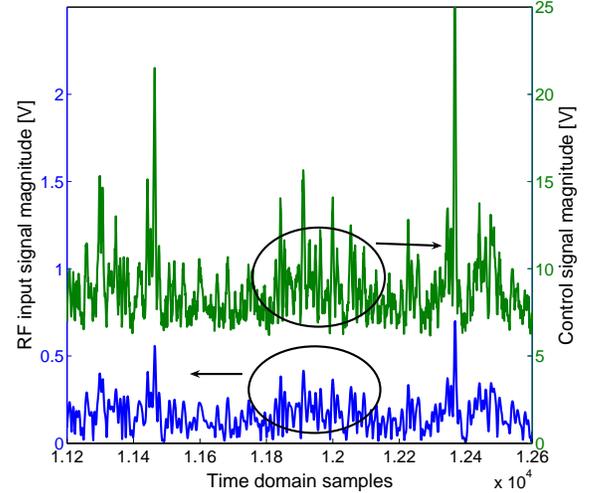}
\caption{Measured magnitude of the RF input signal and the varactor control
signal for the class-E PA.}
\label{vc_inpt}
\end{figure}

The measured magnitude of the complex input signal and the control voltage for the class-E PA are shown in Fig.~\ref{vc_inpt}. From this figure, it can be observed that the baseband control voltage correctly follows the input.

\begin{figure}
\centering
\includegraphics[width=0.9\columnwidth]{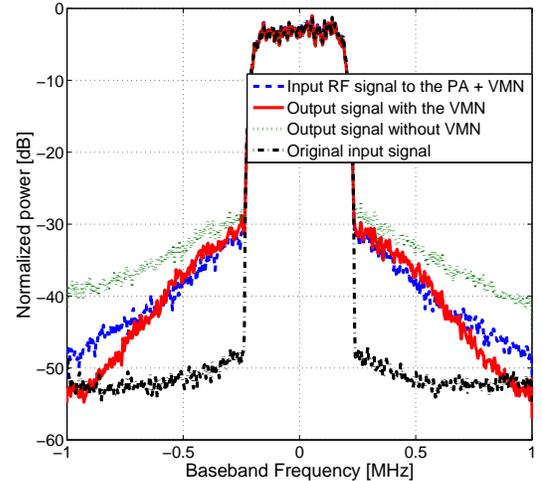}
\caption{Signal spectrum of the RF input signal to the class E PA and output
signal of the VMN.}
\label{spct_rfE}
\end{figure}

Due to severe bandwidth limitations of the class-E PA used, it was not possible to perform the measurements at the full WCDMA bandwidth on this power amplifier. The signal bandwidth was therefore down-scaled with a factor of 10 to 384 kHz for this PA. The measured input and output spectra for the class E PA is shown in Fig.~\ref{spct_rfE}.

It can be noticed that the linearity of the PA+VMN architecture is comparable and slightly better than the linearity of the PA-alone architecture. For a fair comparison, the average output power for both architectures were set to be as close as possible. The output power for the class E PA-alone and PA+VMN architecture were $0.38$ and $0.44$ W respectively.

The PAE measured for the class E PA-alone architecture was measured to be 19\%, while the PA+VMN architecture resulted in a 31\% PAE. The results agree with the theoretical predictions, and a 12\% improvement in PAE for the 11 dB PAR signal is obtained. The ACPR value for both the PA and the PA+VMN is around 27 dB. It can be also be noticed in Fig.~\ref{spct_rfE} that the output signal has less spectral regrowth compared to the PA alone.

\subsection{Results -- Class J PA}
\begin{figure}
\centering
\psfrag{x}[c][c][1]{$x$} \psfrag{y}[c][c][1]{10 W Class-J PA} \psfrag{a}[c][c][1]{Varactor-based} \psfrag{b}[c][c][1]{matching network} \psfrag{z}[c][c][1]{$V_c$} \psfrag{v}[c][c][1]{$y$}
\includegraphics[width=0.9\columnwidth]{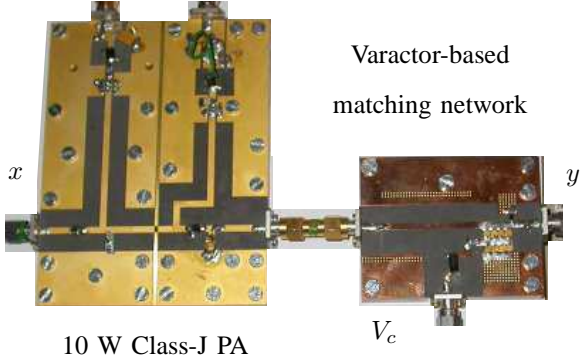}
\caption{Load modulation architecture for the class-J PA.}
\label{classJ}
\end{figure}

In order to test the technique with a full-bandwidth WCDMA signal, a more wideband PA was used. This PA was a class-J LDMOS PA with peak output power of 10 W and operating at 1 GHz \cite{classJ}. A picture of this PA connected to the varactor matching network is shown in Fig.~\ref{classJ}.

\begin{figure}
\centering
\includegraphics[width=0.9\columnwidth]{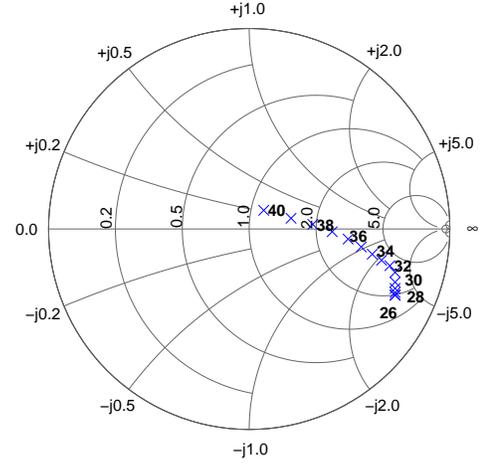}
\caption{Load trajectory for best PAE performance for the class-J PA. The optimum load for each power level is shown.}
\label{classJsmith}
\end{figure}

The class-J PA \cite{classJ} is characterized at 1 GHz by load-pull measurements in a similar way as presented in \cite{hosseinMTT,hosseinEMW}. The optimum load trajectory with the best PAE performance versus output power is then identified and shown in Fig.~\ref{classJsmith}. The load trajectory starts from a point close to the center of Smith Chart and extends out very similar to the class-E PA in Fig.~\ref{classEsmith}. It is, however, rotated in Smith Chart compared to that for the class-E PA. The fact that a similar load trajectory but only rotated is required, implies that the same VMN could be employed for both PAs to control the load impedance optimally. The only thing that has to be done is to adjust the rotation in the Smith Chart. In this work, this is done by replacing the adaptors between the PA and the VMN.

This is an interesting observation indicating that the same VMN design with only delay adjustment can be used for PAs that have the same output power level, operating frequency, and transistor technology. This can be seen as an advantage of the modular design approach proposed in \cite{hosseinMTT} which is experimentally proven in this work.

\begin{figure}
\centering
\psfrag{a}[c][c][1]{RF Input voltage $|x|$ [V]} \psfrag{b}[c][c][1]{RF Output voltage $|y|$ [V]} \psfrag{c}[c][c][1]{Control voltage $V_c$ [V]}
\includegraphics[width=0.85\columnwidth]{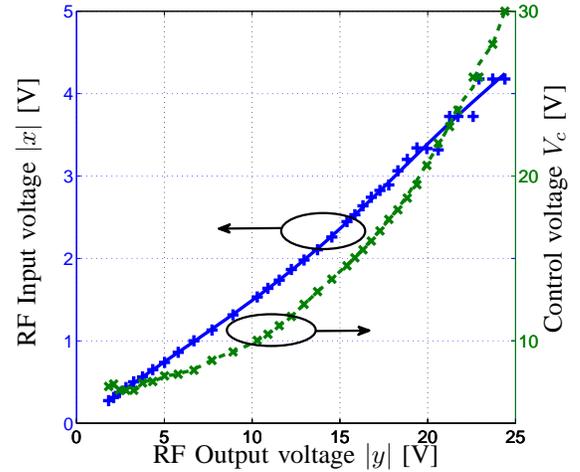}
\caption{Output-input power relations for optimum PAE performance for the class-J PA. The blue crosses are the RF input voltages from static measurements and the blue line is the polynomial approximation. The green pluses are the control voltages from static measurements and the green dashed line is the polynomial approximation.}
\label{out-inJ}
\end{figure}

The input-output transfer function for optimum PAE performance of this PA is shown in Fig.~\ref{out-inJ}. From the static measurement it can be observed that compared to the class E PA, the $|x|$, $V_c$ combination resulting in highest efficiency corresponds to a more linear drive of the input vs. the output.

From equations (\ref{aoutin1})-(\ref{aoutin3}) it can be observed that the phase characteristics do not affect the choice of the optimum $V_c$ and $|x|$. Hence, the phase characteristics need not be known \emph{a priori} and can be obtained directly from the modulated measurements.

\begin{figure}
\centering
\includegraphics[width=0.8\columnwidth]{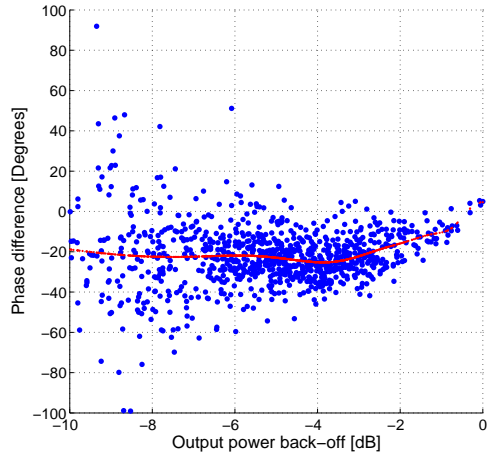}
\caption{Measured phase for the class-J PA + VMN architecture. The blue dots are the initial measured phase and the red line is the polynomial fitted model.}
\label{phaseJ}
\end{figure}

First, the phase difference is calculated from an initial measurement on the architecture between the measured input and output signals. Second, a 7$^{\text{th}}$-order memoryless polynomial model is used to fit the data based on the relationship between the desired output signal and the phase difference. Third, the phase of the predistorted input signals is obtained by using the phase model coefficients as a simple predistorter. The initial measured phase, and the polynomial-fitted model function are shown in Fig.~\ref{phaseJ}.

\begin{figure}
\centering
\includegraphics[width=0.8\columnwidth]{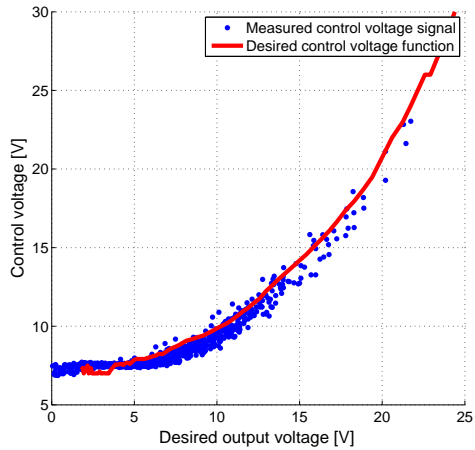}
\caption{The measured control voltage $V_\text{c}$ and the desired control voltage for the VMN+Class J PA.}
\label{vc}
\end{figure}

After following the same procedure as the class E PA, measurements can be done on this PA. In Fig.~\ref{vc}, the measured control voltage signal and the desired control voltage vs the desired output for the class J PA is given. It can be noticed that the signals correspond well, but more advanced modeling is needed to gain more accuracy.

\begin{figure}
\centering
\includegraphics[width=0.8\columnwidth]{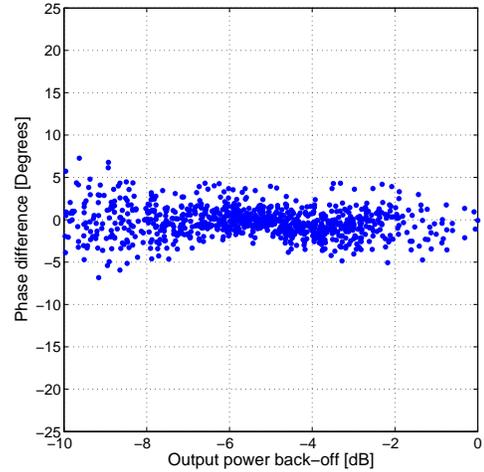}
\caption{Measured phase for the class-J PA + VMN architecture after the pre-distortion.}
\label{phaseJ1}
\end{figure}

The measured output phase difference, after applying the phase pre-distortion, is shown in Fig.~\ref{phaseJ1}. It can be observed that compared to Fig.~\ref{phaseJ} the phase variations are much lower and the phase is almost constant.

\begin{figure}
\centering
\includegraphics[width=0.8\columnwidth]{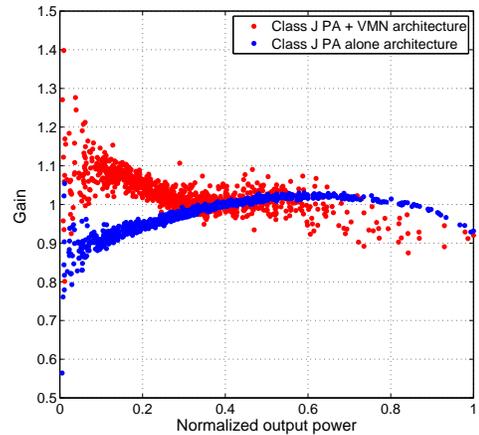}
\caption{A comparison of the normalized gain for the two architectures.}
\label{gainJ}
\end{figure}

The normalized gain for the PA+VMN architecture and the PA-alone architecture is shown in Fig.~\ref{gainJ}. It can be noticed that the PA+VMN architecture provides a more constant gain for the different power levels than the PA-alone architecture.

\begin{figure}
\centering
\includegraphics[width=0.8\columnwidth]{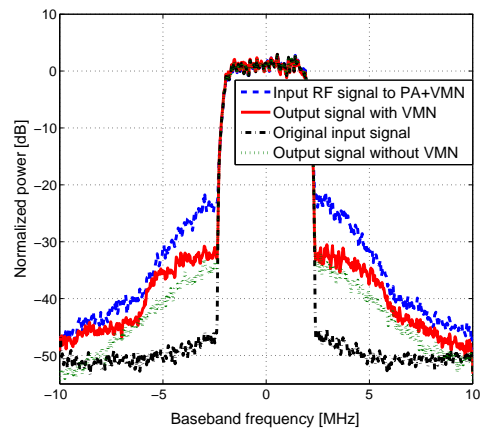}
\caption{Signal spectrum of the full WCDMA bandwidth RF input signal to the class J PA and output signal of the VMN.} \label{spct_rfJ_w}
\end{figure}

The spectrum of the full 3.84 MHz one carrier WCDMA 11.3 dB PAR measurement results for the class J PA is shown in Fig.~\ref{spct_rfJ_w}. It can be observed that the ACPR is around 32 dBc for this signal.

The average PAE was measured to be 28\% for the PA-alone architecture, while the PAE for the PA+VMN architecture was 39\%. Hence, with the help of the VMN, the average PAE is increased by 11\%. The output power for the class J PA with and without the VMN was measured to be 0.59 W and 0.56 W respectively. By using crest factor reduction techniques on this communication signal it would be possible to further improve the power efficiency.

\begin{figure}
\centering
\includegraphics[width=0.8\columnwidth]{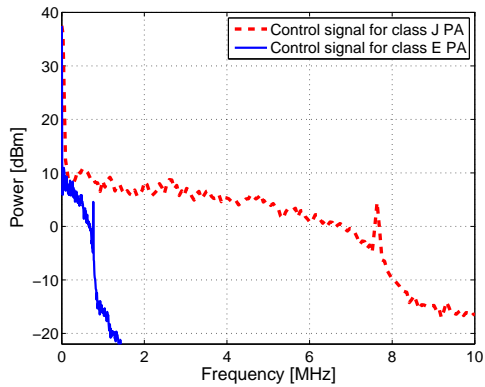}
\caption{Signal spectrum of the baseband input signal to the VMN for the class E and class J power amplifiers. The input signal to the class E PA had a 384 kHz bandwidth, while the class J PA had a 3.84 MHz input signal bandwidth.}
\label{spct_vmn}
\end{figure}

The spectrum for the control voltage for both power amplifiers is also shown in Fig.~\ref{spct_vmn}. It can be noticed from this figure that the signal bandwidths are of practical values. The bandwidth of the control signal for the class E PA which contains $95\%$ of the power is $1.1$ MHz, and for the class J it is 10 MHz. Since the varactor network consumes very little power, the generation of such a signal should not be a challenging issue. It it is worth reminding that the class E PA was tested with a 384 kHz signal, while the class J PA was measured with a 3.84 MHz signal. Therefore, the bandwidth expansion of both PAs relative to the input signal bandwidth is similar.

\subsection{Summary}
From these two experiments, we have effectively shown the practicability of using the same VMN to improve efficiency for two different PAs.

A comparison of the resulting efficiencies is summarized in Table~\ref{results}. It can be noticed that the measured results agree fairly well with the predictions. In this table we have also included results from measurement on the class J PA with a reduced bandwidth similar to the class E PA experiment.

\begin{table}
%\centering \caption{Efficiency results for the modulated
%measurements}
%\begin{tabular}{c|c}
%   % after \\: \hline or \cline{col1-col2} \cline{col3-col4} ...
%   Architecture  & Average PAE \%\\
%   \hline
%   \hline
%   Class E PA from static estimation & 21 \\
%   Class E PA + VMN from static estimation & 30 \\
%   Class E PA with measurements & 19 \\
%   Class E PA + VMN with measurements & 31\\
%   \hline
%   Class J PA from static estimation & 27\\
%   Class J PA + VMN from static estimation & 41\\
%   &\\
%   Class J PA with measurements &\\
%   Reduced bandwidth&31\\
%   Class J PA + VMN with measurements &\\
%   Reduced bandwidth&41\\
%   \hline
%   Class J PA with measurements &\\
%   Full bandwidth&28\\
%   Class J PA + VMN with measurements &\\
%   Full bandwidth&39
%   \end{tabular}
%   \label{results}
%\end{table}
\centering
\caption{Efficiency results for the modulated measurements}
\begin{tabular}{c|c|c}

   % after \\: \hline or \cline{col1-col2} \cline{col3-col4} ...
   & \multicolumn{2}{c}{Average PAE \%}\\
   Architecture&from static& with\\
   &estimation&measurements\\
   \hline
   \hline
   Class E PA&21&19\\
   Class E PA +VMN & 30&31\\
   \hline
   Class J PA reduced bandwidth&27&31\\
   Class J PA + VMN reduced bandwidth&41&41\\
   \hline
   Class J PA full bandwidth&27&28\\
   Class J PA + VMN full bandwidth&41&39\\
   \end{tabular}
   \label{results}
\end{table}

With the help of the VMN network, the power efficiency of two available high efficiency power amplifiers was enhanced by 10-14\% for modulated signals. As the need for better spectral efficiency grows and correspondingly the PAPR of communication signals grows, such transmitter architectures can help increase the power efficiency of communication system.

\section{Conclusion}

In this work we have demonstrated a new architecture with modulated measurements. The feasibility of dynamic load modulation with varactor based networks for two high power high efficiency amplifiers was shown in this work. A simple and efficient dual input quasi-static linearization scheme, which is also useful for other dual input architectures, such as DSM was demonstrated. It was also shown that due to the modular design of the architecture, the same varactor matching network could be used with two different PAs.

The results shows significant improvement in back-off and average power efficiency for the architecture compared to a normal PA configuration with a fixed $50~\Omega$ load impedance can be achieved. With a careful co-design of the input signals, up to 14 percentage units average efficiency improvement was achieved. The dual-input nature of the DLM architecture has great modeling challenges, but also gives great possibilities for further improving the linearity and efficiency.

% use section* for acknowledgement

\bibliographystyle{IEEEtran}
\bibliography{main}

\end{document}